**Partition-based Field Normalization: an Approach to Highly Specialized Publication Records.**

**Nadine Rons**

Research Coordination Unit and Centre for R&D Monitoring (ECOOM), Vrije Universiteit Brussel (VUB), Pleinlaan 2, B-1050 Brussels, Belgium; *E-mail address:* Nadine.Rons@vub.ac.be

*Telephone number:* +32-2-629 21 79 or +32-2-629 10 09 (secr.)

*Fax number:* +32-2-629 36 40

**ABSTRACT**

Field normalized citation rates are well-established indicators for research performance from the broadest aggregation levels such as countries, down to institutes and research teams. When applied to still more specialized publication sets at the level of individual scientists, also a more accurate delimitation is required of the reference domain that provides the expectations to which a performance is compared. This necessity for sharper accuracy challenges standard methodology based on pre-defined subject categories. This paper proposes a way to define a reference domain that is more strongly delimited than in standard methodology, by building it up out of cells of the partition created by the pre-defined subject categories and their intersections. This partition approach can be applied to different existing field normalization variants. The resulting reference domain lies between those generated by standard field normalization and journal normalization. Examples based on fictive and real publication records illustrate how the potential impact on results can exceed or be smaller than the effect of other currently debated normalization variants, depending on the case studied. The proposed *Partition-based Field Normalization* is expected to offer advantages in particular at the level of individual scientists and other very specific publication records, such as publication output from interdisciplinary research.

## 1. Introduction

The field normalized citation rate, introduced by Braun and Glänzel (1990) and Moed et al. (1995), is considered among the most important bibliometric indicators for research performance. This indicator is claimed to correlate well in general with peer judgments (Van Raan, 2006a and 2006b). Nevertheless, correlations vary depending on factors such as research domain or evaluation methodology (Aksnes and Taxt, 2004; Rons and Spruyt, 2006), and studies comparing bibliometric results to peer judgments, offering a two-way reliability test, are limited in number and scope. In some particular circumstances, the appropriateness of the field normalized citation rate is questioned since its launch, i.e. for multidisciplinary oriented research groups and where the definition of the research domain in terms of journal categories is inadequate (Moed et al., 1995, p. 404 and 410). Such domain definition problems are encountered when the indicator is applied to highly specialized publication records at the level of individual scientists. A reference domain needs to be determined that accurately represents a relatively small research community working in a same research area, to provide appropriate expected citation rates and produce adequately normalized results. In the current research policy context that increasingly focuses on individual excellence, this problem figures among the important challenges to standard calculation methods for a field normalized citation rate. Some alternative indicators designed specifically for individual researchers do not use a fixed subject category structure in their calculations, e.g. focusing on the evolution of citation impact rather than on its value at a particular moment in Impact Vitality (Rons & Amez, 2009), or on the accumulated volume of most cited papers in the H-Index (Hirsch, 2005). Another way to create a sufficiently specific reference domain for the work of specialized research groups and individual scientists, is to use paper-by-paper field delimitation methods (Schubert & Braun, 1996). The resulting reference standard can be more specialized than is possible with journal-based methods, with a decisive effect on performance evaluations based on the normalized citation rates (Bornmann et al., 2008). Such paper-based classification schemes are available in several specialized databases dedicated to a specific research area, yet not in the most general and widely used databases for bibliometric analysis, Web of Science and Scopus, structured in fixed journal-based subject categories. This paper proposes an adaptation of standard field normalization methodology that still starts from a fixed subject category structure, but that more strongly delimits the reference domain around an observed publication record. Instead of the relatively large entire subject categories, *Partition-based Field Normalization* uses as building blocks the smaller cells of the partition created by these subject categories and their intersections. This partition approach can be combined with different existing field normalization variants that are currently mainly based on the highly overlapping journal-based subject categories of Web of Science or Scopus. The paper-by-paper field delimitation methods are not a main area of application for the partition approach, as these can delimit very specific fields consisting of parts of journals, and are less subject to overlaps because only the topic of the paper itself determines its classification. In these perspectives of usage and expected usefulness, the paper further focuses on an application of the partition approach to standard field normalization variants in an environment of journal-based subject categories.

## 2. Methodology

### 2.1. Field normalization variants

The novel element introduced by Partition-based Field Normalization with respect to standard methodology, lies in the way that it treats publications in intersections of subject categories, i.e. publications in journals classified in more than one subject category. Partition-based Field Normalization bases expected citation rates for publications in such intersections only on the journals in that particular intersection. Before describing the partition approach in more detail, this section briefly discusses how established field normalization variants handle publications in intersections of subject categories, and what it implies.

In current standard calculations of the field normalized citation rate, a paper in a journal classified in $N$ subject categories is counted as $1/N$ paper in each subject category. This is applied in its different variants: the Citations Per Publication normalized with respect to the mean Field Citation Score or *CPP/FCSm* (Moed et al., 1995), the similar Normalized Mean Citation Rate *NMCR* (Braun & Glänzel, 1990) and the Mean Normalized Citation Rate MNCR that will replace *CPP/FCSm* as the new 'Crown Indicator' (Lundberg, 2007; Waltman et al., 2011a). These field normalization variants differ in the level at which citations are normalized: per publication for *MNCR*, and at global level for *CPP/FCSm* and *NMCR*. For the remaining part they are very similar, including two key choices in common determining the expected citation rate for publications in an intersection of subject categories, listed in Table 1. Together with the particular mean value calculated over the expected citation rates for the different intersecting subject categories (the harmonic mean for *MNCR*, and the arithmetic mean for *CPP/FCSm*), these choices result in the property that the indicator has a value of one when calculated for the all-encompassing set of all publications published in all subject categories. The value of one is therefore interpreted as a world average.

**Table 1.** Choices in standard field normalization determining the expected citation rate for publications in an intersection of subject categories.

| Choice | Adequate for: | Less appropriate for: |
|---|---|---|
| 1. The expected citation rate is determined by the characteristics of all **entire** intersecting subject categories, i.e. including their parts outside the intersection. | Publication sets of large entities, spread out over a wide range of subject categories. | Publication sets of small entities in very specialized research areas, including publications from only few subject categories. |
| 2. **Equal** weights $1/N$ are given to each of the $N$ intersecting subject categories. | Intersections where citation practices lie between those associated with each of the intersecting subject categories separately. | Intersections where citation practices lie outside those associated with each of the intersecting subject categories separately. |

Depending on circumstances, both choices mentioned above can be more or less adequate. Table 1 lists in which way the adequacy can be expected to depend on (1) the volume of the observed publication set and (2) the intermediacy of intersection characteristics with respect to the intersecting subject categories.
(1) For a publication record including publications in an intersection of subject categories, some of these subject categories may for their main part (outside the intersection) be unrelated to the observed record's research domain, adjacent to but outside its border. Yet in standard field normalization, also these adjacent parts of subject categories will contribute to the calculation of the expected citation rate for publications in an intersection. For smaller publication records, with a relatively larger border region, such unrelated parts may influence the expected citation rates for a larger fraction of its publications.

(2) Citation practices characteristic for publications in an intersection of subject categories may not be situated in between those that are characteristic for each of the intersecting subject categories separately. In such cases standard field normalization nevertheless assumes that they are, using an average over the intersecting subject categories with equal weights to calculate the expected citation rate for publications in an intersection.

In both cases, standard field normalization may include unrelated citation practices into the calculations, instead of controlling for differences in citation potential among research fields, which is the purpose of field normalization.

Given the ample occurrence of intersections of subject areas in the fixed subject category structures of the Web of Science and Scopus, the most general and widely used databases for bibliometric analysis, the above mentioned choices will be less appropriate for the analysis of numerous cases of highly specialized publication records. Boyack et al. (2005) found an average of 1,59 subject categories associated per journal in the combined Science Citation Index (SCI) and Social Sciences Citation Index (SSCI) in an analysis showing that the Web of Science subject categories in many cases do not reflect current journal groupings based on similarity measures. Table 2 gives an example of an intersection of two subject categories in the Web of Science, illustrating how research communities and citation practices can differ between the intersection of two subject categories on the one hand, and the main parts of the intersecting subject categories outside the intersection on the other hand:

(1) The research community publishing in a particular intersection of subject categories may to a large extent be different from the research communities publishing in the parts of these subject categories outside this intersection (in the example in Table 2, 58% of authors publishing in the intersection are not in common with either of the two subject categories excluding the intersection).

(2) The impact factors of journals in a particular intersection of subject categories, and therefore the expected citation rate calculated from the whole set of journals in that intersection (aggregate impact factor), can lie well outside the range of the expected citation rates for the intersecting subject categories (compare the value 2,659 for the intersection to the values 1,649 and 1,331 for the two intersecting subject categories separately in the example in Table 2).

Standard calculations of expected citation rates for publications in intersections include contributions from such possibly unrelated research communities and their citation practices. The next section describes how this is avoided by the proposed partition approach.

**Table 2.** Example of an intersection of two subject categories.

| | Subject Category 1: COMPUTER SCIENCE, INTERDISCIPLINARY APPLICATIONS | | INTERSECTION[a] | Subject Category 2: INFORMATION SCIENCE & LIBRARY SCIENCE | |
|---|---|---|---|---|---|
| | $S_1$ (incl. $I_{1,2}$) | $S_{1e}$ (excl. $I_{1,2}$) | $I_{1,2}$ = $S_1 \cap S_2$ | $S_{2e}$ (excl. $I_{1,2}$) | $S_2$ (incl. $I_{1,2}$) |
| Number of journals [b] | 95 | 92 | 3 | 63 | 66 |
| Number of articles and reviews [b] | 9576 | 9246 | 330 | 2289 | 2619 |
| Number of articles and reviews per journal | 100,8 | 100,5 | 110,0 | 36,3 | 39,7 |
| **Impact factors** [c] | | | | | |
| - Aggregate | **1,649** | 1,618 | **2,659** | 1,191 | **1,331** |
| - Maximum | 3,974 | 3,882 | 3,974 | 4,485 | 4,485 |
| - Minimum | 0,203 | 0,203 | 0,635 | 0,000 | 0,000 |
| **% of authors [d] publishing in $I_{1,2}$ that are:** | | | | | |
| - in common with authors [d] publishing in $S_{1e}$, but not in $S_{2e}$ | | | 17,4% | | |
| - in common with authors [d] publishing in $S_{2e}$, but not in $S_{1e}$ | | | 7,5% | | |
| - in common with authors [d] publishing in both $S_{1e}$ and $S_{2e}$ | | | 16,4% | | |
| - not in common with authors [d] publishing in $S_{1e}$ or $S_{2e}$ | | | **58,6%** | | |

[a] Consisting of the journals SCIENTOMETRICS, JOURNAL OF THE AMERICAN MEDICAL INFORMATICS ASSOCIATION, and SOCIAL SCIENCE COMPUTER REVIEW.
[b] Source: Journal Citation Reports (JCR) 2009, last accessed online 13.05.2011.
[c] (Cites in 2009 to items in 2008 and 2007) / (Number of items published in 2008 and 2007). Source: JCR 2009, last accessed online 13.05.2011.
[d] Distinct name-and-initials combinations as authors of publications (all document types) in index years 2005 to 2009. Source: Web of Science (WoS), last accessed online 19.05.2011.
*Data sourced from Thomson Reuters Web of Knowledge (formerly referred to as ISI Web of Science).*

## 2.2. Partition-based Field Normalization

The above discussion and example of an intersection of subject categories indicate that, for a highly specialized research record, the appropriate reference domain and expected citation rate for publications in an intersection may be more accurately determined when based only on the journals in that intersection. This can be implemented in a straightforward way in practice, by calculating expected citation rates not per original subject category, but per cell of the partition formed by the fixed subject categories and their intersections. The set *X* of all publication sources is divided in a set of non-empty subsets of *X* such that every publication source in *X* is in exactly one of these subsets, being the subset that contains all publication sources classified in exactly the same combination of subject categories. These non-overlapping and non-empty subsets will in this paper be called "cells" of the partition of *X*. The union of all cells covers *X* and the intersection of any two distinct cells is empty. Subsequently, the 'unidisciplinary' cells and 'pluridisciplinary' cells are treated in a same way, requiring no further weights or choices.

Table 3 compares how citations are normalized, on the one hand in standard field normalization and on the other hand in Partition-based Field Normalization, based on a same fixed subject category structure. The partition approach avoids expected citation rates to be influenced by citation practices typical for cells situated close to the cells that contain publications from an observed record, but which are not representing its research domain. Compared to standard methodology, Partition-based Field Normalization leaves those cells out of scope that themselves do not contain elements from the observed publication record, but which are part of a larger subject category that *does* contain some, in one or more of the other cells that are part of that subject category.

**Table 3.** A comparison of reference domains in standard field normalization versus Partition-based Field Normalization.

| Location of an observed publication | Journals included in the reference domains generating the expected citation rates | |
|---|---|---|
| | **Standard Field Normalization** | **Partition-based Field Normalization** |
| A journal classified in one subject category only | All journals classified in that subject category, with partial contributions 1/$N$ from journals classified in $N$ subject categories. | All journals in the cell containing all journals classified in that subject category only. |
| A journal classified in a combination of multiple subject categories | All journals classified in any of the combined subject categories, with equal weights given to each subject category, and per subject category with partial contributions 1/$N$ from journals classified in $N$ subject categories. | All journals in the cell containing all journals classified in exactly the same combination of subject categories. |

This difference between Partition-based Field Normalization and the standard approach is visualized in Figure 1 for a fictive publication record containing publications in four of the seven cells formed by four intersecting subject categories. Two of the three cells not containing elements from the publication record are left out of scope in Partition-based Field Normalization, but would be included in the calculations in standard methodology (shaded).

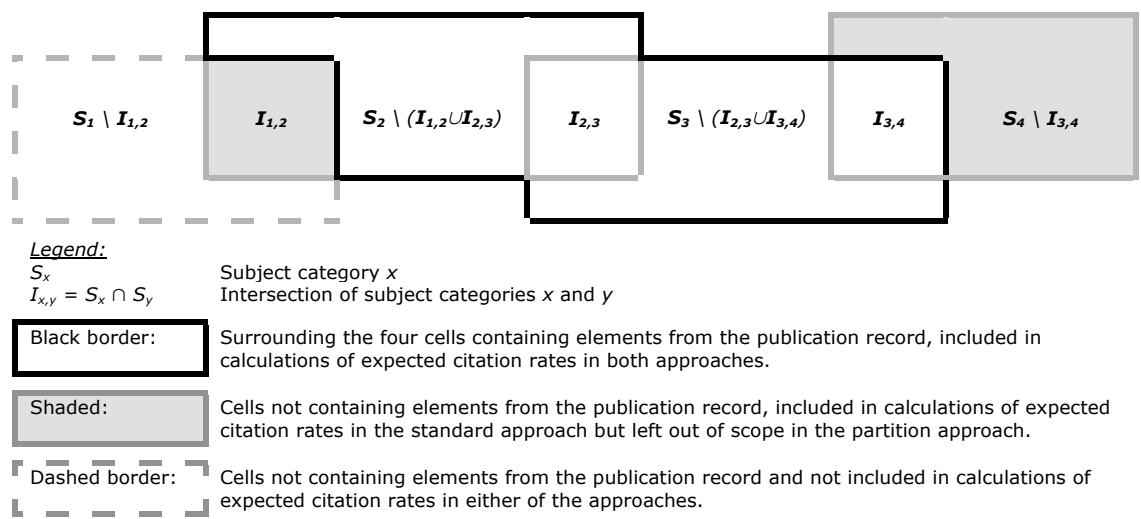

Legend:
$S_x$  Subject category $x$
$I_{x,y} = S_x \cap S_y$  Intersection of subject categories $x$ and $y$

Black border: Surrounding the four cells containing elements from the publication record, included in calculations of expected citation rates in both approaches.

Shaded: Cells not containing elements from the publication record, included in calculations of expected citation rates in the standard approach but left out of scope in the partition approach.

Dashed border: Cells not containing elements from the publication record and not included in calculations of expected citation rates in either of the approaches.

**Figure 1**. Partition-based Field Normalization: Example of cells included and cells left out of scope with respect to standard methodology.

In this way, Partition-based Field Normalization defines a more strongly delimited set of journals as the global reference domain that is to describe an observed publication record's research area, and the corresponding research community and its citation behavior.

**3. Amplitude of effects**

The factors that, from the discussions in the preceding sections, can be expected to increase the effect of the partition approach on the field normalized citation rate for an observed publication record, are summarized in Table 4.

**Table 4.** Factors increasing the effect of the partition approach on the field normalized citation rate.

| Factor | Rationale |
|---|---|
| (1) A more strongly specialized research area. | Small research areas are more likely to give rise to a high relative volume of cells in the border area that is left out of scope in the partition approach compared to the standard approach, affecting expected citation rates for a high fraction of publications. |
| (2) A larger fraction of publications published in journals classified in multiple subject categories. | Intersections of subject categories are the primary locations of the difference between the standard and the partition approach. |
| (3) Larger differences in expected citation rates between intersecting subject categories. | When more strongly differing citation practices are left out of scope in the partition approach compared to the standard approach, their respective results will more strongly differ. |
| (4) Intersections of subject categories with characteristics situated outside those of the intersecting subject categories separately. | For publications in such an intersection, the expected citation rate in the partition approach will more strongly differ from the one in the standard approach determined as an average over the intersecting subject categories with equal weights. |

Small publication records will more easily be subject to these factors or more sensitive to their effects, than publication samples at a higher aggregation level. The effect of Partition-based Field Normalization can therefore be expected to be high at the level of individual scientists, and their comparison. Exactly at this level of individual scientists, standard field normalization is not applied with a same level of confidence as is generally attached to analyses at the level of teams and larger entities. The partition approach may thus offer an improvement in particular at this level.

Below, two examples illustrate how applying the partition approach can influence relative results between two publication sets of limited size, in one example based on fictive and in the other on real publication records. Field normalized citation rates are calculated in four variants, distinguishing on the one hand between global normalization (*NMCR*) or normalization per publication (*MNCR*), and on the other hand between a standard approach and application of the partition approach, where the latter will in further notation be indicated by the prefix "*P-*" (*P-NMCR* and *P-MNCR*).

### 3.1. First example based on fictive publication records

Observe as a fictive example two extremely simplified publication records, situated in the realistic context provided by the intersection of two subject categories illustrated in Table 2. All publications in both records belong to Subject Category 2, and partly to the intersection with Subject Category 1. The records differ in their ratios of publications belonging to Subject Category 2 exclusively $S_{2e}$, versus those belonging to the intersection $I_{1,2}$, which are inversed. Both ratios are 'realistic' in the sense that they may occur in a real scientist's research record. For this demonstration, the following simplifying assumptions are added:

(1) Each publication yields exactly *A* times the expected citations calculated for the partition cell in which it is situated.
(2) The observation is limited to the year 2009 only, with expected citation rates *E* equal to the aggregate impact factors, per cell in Partition-based Field Normalization ($E_{1e}$=1,618, $E_I$=2,659, $E_{2e}$=1,191), and per entire subject category in standard methodology ($E_1$=1,633, $E_2$=1,265), where indices indicate the intersection of the two subject categories ($_I$) and both subject categories either as a whole ($_1$, $_2$) or excluding the intersection ($_{1e}$, $_{2e}$), and where publications in the intersection are counted for 1/2 in $E_1$ and $E_2$.

Table 5 shows the field normalized citation rates in the four different variants. This example shows that applying the partition approach can have a significant effect on the relative values *Q* of normalized citation rates for different publication records in a sample (difference of about one quarter between $Q_a$ and $Q_b$ for *P-NMCR* and *NMCR*, and between $Q_c$ and $Q_d$ for *P-MNCR* and *MNCR*). In this particular example, the effect of applying the partition approach strongly exceeds the (almost) neutral effect of switching between global normalization and normalization per publication ($Q_a = Q_c$, and $Q_b$ lies very close to $Q_d$).

**Table 5**

First example: two fictive publication records in a realistic setting.

| Composition (cf. Table 2) | Publication record 1 | Publication record 2 | |
|---|---|---|---|
| -Share in $I_{1,2}$ | 2/3 | 1/3 | |
| -Share in $S_{2e}$ | 1/3 | 2/3 | |
| Normalization variant | Field normalized citation rate ($R$) | | Ratio |
| | $R_1$ | $R_2$ | $Q=R_1/R_2$ |
| (a) *P-NMCR* | $(2*A*E_I+A*E_{2e})/(2*E_I+E_{2e})=A$ | $(A*E_I+2*A*E_{2e})/(E_I+2*E_{2e})=A$ | $Q_a$=1.00 |
| (b) *NMCR* | $(2*A*E_I+A*E_{2e})/(2*(E_1+E_2)/2+E_2)$ $=A*(2*E_I+E_{2e})/(E_1+2*E_2)=1.564*A$ | $(A*E_I+2*A*E_{2e})/((E_1+E_2)/2+2*E_2)$ $=A*(E_I+2*E_{2e})/(E_1/2+5*E_2/2)=1.267*A$ | $Q_b$=1.23 |
| (c) *P-MNCR* | $(2*A*E_I/E_I+A*E_{2e}/E_{2e})/3=A$ | $(A*E_I/E_I+2*A*E_{2e}/E_{2e})/3=A$ | $Q_c$=1.00 |
| (d) *MNCR* | $(2*A*E_I*(1/E_1+1/E_2)/2+A*E_{2e}/E_2)/3=1.557*A$ | $(A*E_I*(1/E_1+1/E_2)/2+2*A*E_{2e}/E_2)/3=1.250*A$ | $Q_d$=1.25 |

### 3.2. Second example based on real publication records

In this second example, ten real publication records are compared, for demonstration purposes limited to the five most cited articles in a five-year period (Table 6). The subjects are ten candidates for a research fellowship, with different sub-specializations within a same research area. Depending on the sub-specialization, their most cited articles are published in a different set of journals, situated in a different combination of partition cells. A normalization of citations fit to the candidates' sub-specializations may therefore be needed to generate normalized results that allow an accurate comparison between candidates in different sub-specializations. If Partition-based Field Normalization can provide a higher accuracy to this respect than standard field normalization variants, then its indicator values should be better correlated with peer ratings, where expectations for the sub-specialization are taken into account based on the peers' expertise.

**Table 6.** Second example: Most cited articles of ten researchers in a same research area.

| Researcher | 5 most cited articles from 2004-2008, ranked I to V | | | | | | | | | | | | | | |
|---|---|---|---|---|---|---|---|---|---|---|---|---|---|---|---|
| | Journal | | | | | Publication year | | | | | Number of citations until 2010 | | | | |
| | I | II | III | IV | V | I | II | III | IV | V | I | II | III | IV | V |
| Researcher 1 | $J_1$ | $J_2$ | $J_3$ | $J_2$ | $J_2$ | '06 | '04 | '04 | '08 | '07 | 226 | 180 | 125 | 74 | 71 |
| Researcher 2 | $J_4$ | $J_2$ | $J_5$ | $J_2$ | $J_6$ | '04 | '05 | '05 | '04 | '07 | 298 | 278 | 133 | 86 | 40 |
| Researcher 3 | $J_1$ | $J_2$ | $J_2$ | $J_2$ | $J_2$ | '06 | '04 | '08 | '07 | '07 | 226 | 180 | 74 | 71 | 59 |
| Researcher 4 | $J_1$ | $J_2$ | $J_2$ | $J_2$ | $J_2$ | '06 | '04 | '04 | '05 | '05 | 226 | 180 | 58 | 54 | 36 |
| Researcher 5 | $J_7$ | $J_1$ | $J_8$ | $J_9$ | $J_{10}$ | '06 | '05 | '05 | '07 | '08 | 9 | 2 | 1 | 0 | 0 |
| Researcher 6 | $J_2$ | $J_4$ | $J_7$ | $J_7$ | $J_7$ | '04 | '04 | '05 | '06 | '06 | 276 | 136 | 69 | 66 | 64 |
| Researcher 7 | $J_4$ | $J_7$ | $J_7$ | $J_7$ | $J_7$ | '04 | '05 | '06 | '06 | '05 | 136 | 69 | 66 | 64 | 63 |
| Researcher 8 | $J_{11}$ | $J_2$ | $J_2$ | $J_{11}$ | $J_2$ | '08 | '08 | '08 | '08 | '08 | 144 | 139 | 96 | 63 | 50 |
| Researcher 9 | $J_5$ | $J_2$ | $J_2$ | $J_5$ | $J_2$ | '05 | '04 | '06 | '05 | '08 | 329 | 249 | 170 | 125 | 96 |
| Researcher 10 | $J_7$ | $J_7$ | $J_7$ | $J_4$ | $J_4$ | '07 | '04 | '04 | '05 | '05 | 51 | 48 | 24 | 23 | 14 |

Source: Web of Science (WoS), last accessed online 10.09.2011.
*Data sourced from Thomson Reuters Web of Knowledge (formerly referred to as ISI Web of Science).*

For demonstration purposes a limited publication universe is observed, consisting of the journals in the cells containing these most cited articles, completed with all other journals from each of the subject categories involved that are classified in that subject category only (Source: JCR 2009). For the simplicity of the example and easy reproducibility of calculations, all more peripheral cells were disregarded. The resulting set of journals and associated subject categories is listed in Table 7. All journals involved in the calculation of the Partition-based Field Normalization variants are included. From the additional journals used in the calculations of the standard field normalization variants, the most peripheral ones are excluded. In this way, the standard field normalization variants calculated in this simplified example are de facto approximated, already using a more specialized reference domain than would be the case in an unlimited publication universe.

**Table 7.** Second example: Article and citation data needed to calculate expected citation rates.

| Journal | Subject categories the journal is classified in | Per publication year: (Number of articles published, Number of citations received until 2010) | | | | |
|---|---|---|---|---|---|---|
| | | 2004 | 2005 | 2006 | 2007 | 2008 |
| *Journals in the cells containing the candidates' most cited articles (Table 6):* | | | | | | |
| $J_1$ | $S_1, S_2, S_3, S_4$ | (1721, 11048) | (1355, 8990) | (1643, 7259) | (1886, 6541) | (871, 2434) |
| $J_2$ | $S_5$ | (3575, 145480) | (3692, 125678) | (3758, 105279) | (3545, 78090) | (3905, 64803) |
| $J_3$ | $S_6$ | (916, 136688) | | | | |
| $J_4$ | $S_5$ | (1036, 22966) | (953, 18003) | (997, 15271) | (837, 11505) | (918, 9174) |
| $J_5$ | $S_2, S_7$ | (2206, 51808) | (2161, 44028) | (2287, 39834) | (2177, 32220) | (2750, 29020) |
| $J_6$ | $S_2, S_7$ | (131, 2916) | (147, 3145) | (203, 3923) | (262, 3691) | (321, 3600) |
| $J_7$ | $S_2$ | (649, 4730) | (342, 4180) | (270, 2531) | (359, 2557) | (314, 1853) |
| $J_8$ | $S_5$ | (547, 1125) | (299, 587) | (570, 817) | (0, 0) | (0, 0) |
| $J_9$ | $S_5$ | (246, 1128) | (309, 1149) | (298, 734) | (363, 903) | (274, 299) |
| $J_{10}$ | $S_1$ | (0, 0) | (0, 0) | (43, 136) | (79, 198) | (93, 468) |
| $J_{11}$ | $S_2, S_7$ | (103, 2138) | (90, 1758) | (87, 1101) | (107, 1263) | (88, 782) |
| *All other journals from each of the subject categories involved that are classified in that subject category only (JCR 2009):* | | | | | | |
| | $S_1$ | (101, 701) | (161, 836) | (166, 843) | (243, 1887) | (250, 1846) |
| | $S_2$ | (1357, 27149) | (1380, 24556) | (1407, 20283) | (1523, 18938) | (1589, 17527) |
| | $S_3$ | (1535, 13058) | (1797, 15080) | (1400, 9487) | (1717, 9051) | (1675, 6487) |
| | $S_4$ | (1500, 5467) | (2184, 6953) | (2024, 6154) | (2301, 4986) | (2827, 4020) |
| | $S_5$ | (8450, 49897) | (9025, 51516) | (9376, 50736) | (11276, 48096) | (12839, 43789) |
| | $S_6$ | (7418, 366766) | | | | |
| | $S_7$ | (8324, 194774) | (8345, 163913) | (9095, 156132) | (9184, 121031) | (8309, 71890) |

Source: Web of Science (WoS), last accessed online 14.09.2011.
*Data sourced from Thomson Reuters Web of Knowledge (formerly referred to as ISI Web of Science).*

Table 8 shows the field normalized citation rates in the four different variants for all candidates, as well as selected ratings by peers from the program's evaluation procedure. The peer ratings selected for this comparison are the global average over the ratings on all aspects, and the rating on the 'Scientific background of the applicant', which can be expected to be related to the citation impact of the publications. Normalized citation rates and peer ratings are compared between two individual candidates on the one hand (1), and for the total sample of ten candidates on the other hand (2).

(1) *Results for the two candidates shortlisted in the peer review process:* All normalization variants give rise to a clear distinction between the two candidates, yet in inverse directions for global normalization versus normalization per publication: in favor of researcher 2 for *P-NMCR* and *NMCR* ($Q_a$ and $Q_b$ < 1), and in favor of researcher 1 for *P-MNCR* and *MNCR* ($Q_c$ and $Q_d$ > 1). These strongly differing results for global normalization versus normalization per publication are remarkable yet not surprising, as empirical analysis already revealed larger differences at lower aggregation levels such as research groups and journals (Waltman et al., 2011b). In this particular example, the effect of applying the partition approach (difference between $Q_a$ and $Q_b$, and between $Q_c$ and $Q_d$) is smaller than the effect of switching between global normalization and normalization per publication (difference between $Q_a$ and $Q_c$, and between $Q_b$ and $Q_d$). These results for two individual researchers show that, apart from the factors mentioned above (Table 4), the effect of the partition approach can depend on the normalization variant that it is applied to: here strengthening the distinction between the two researchers with normalization per publication ($Q_c > Q_d > 1$) and weakening it with global normalization ($Q_b < Q_a < 1$).

(2) *Results for the total sample of candidates:* Table 8 shows both Pearson correlations (measuring to what degree two indicators are linearly related) and Spearman correlations (measuring to what degree two indicators yield a same ranking of items) for all indicator pairs of peer ratings and normalized citation rates. Correlations are significant ($p≤0,05$) for only a minority of the indicator pairs. The strongest correlations are reached between peer ratings on 'Scientific background of the applicant' and a combination of the partition approach with normalization per publication (*P-MNCR, p=0,03*). Correlations improve in general with application of the partition approach. This improvement may be underestimated considering that to simplify this example the most peripheral cells were disregarded, which leaves publication practices from the most remote areas out of the standard calculations. These results for the total sample of applicants show that Partition-based Field Normalization can make a step towards more significant correlations with peer ratings.

**Table 8.** Second example: Comparison of field normalization variants and peer ratings.

| | Field Normalized Citation Rate ($R$) | | | | Average peer ratings (range 1-5, best score = 1) | |
|---|---|---|---|---|---|---|
| **Candidates** ($N$=10) | **a) *P-NMCR*** | **b) *NMCR*** | **c) *P-MNCR*** | **d) *MNCR*** | **Scientific background of applicant** | **All aspects** |
| *Shortlisted candidates:* | | | | | | |
| Researcher 1 | 7,04 | 6,86 | 16,80 | 15,15 | 1,67 | 1,44 |
| Researcher 2 | 10,35 | 10,92 | 10,71 | 10,96 | 2,00 | 1,70 |
| Ratio $Q=R_1/R_2$ | $Q_a$=0,68 | $Q_b$=0,63 | $Q_c$=1,57 | $Q_d$=1,38 | | |
| Researcher 3 | 13,79 | 13,04 | 17,75 | 16,10 | 2,00 | 1,70 |
| Researcher 4 | 8,68 | 8,34 | 14,52 | 12,87 | 2,00 | 1,78 |
| Researcher 5 | 0,24 | 0,24 | 0,21 | 0,21 | 2,67 | 1,96 |
| Researcher 6 | 8,07 | 7,96 | 7,91 | 7,85 | 2,33 | 2,04 |
| Researcher 7 | 5,20 | 5,11 | 5,20 | 5,13 | 2,00 | 2,06 |
| Researcher 8 | 12,03 | 12,74 | 12,57 | 13,06 | 2,67 | 2,30 |
| Researcher 9 | 12,95 | 13,69 | 13,44 | 13,96 | 2,50 | 2,33 |
| Researcher 10 | 2,26 | 2,17 | 2,34 | 2,26 | 4,00 | 2,44 |
| **Average peer ratings** | **Pearson product-moment correlations** | | | | | |
| Scientific background of applicant | $r$=-0,43 $p$=0,11 | $r$=-0,40 $p$=0,13 | $r$=-0,61 **$p$=0,03** | $r$=-0,59 **$p$=0,04** | | |
| All aspects | $r$=-0,12 $p$=0,37 | $r$=-0,07 $p$=0,43 | $r$=-0,50 $p$=0,07 | $r$=-0,42 $p$=0,11 | | |
| **Average peer ratings** | **Spearman's rank-order correlations** | | | | | |
| Scientific background of applicant | $r$=-0,23 $p$=0,27 | $r$=-0,18 $p$=0,31 | $r$=-0,63 **$p$=0,03** | $r$=-0,52 $p$=0,06 | | |
| All aspects | $r$=-0,13 $p$=0,36 | $r$=-0,05 $p$=0,44 | $r$=-0,52 $p$=0,06 | $r$=-0,41 $p$=0,12 | | |

Source of underlying bibliometric data (cf. Tables 6 and 7): Web of Science (WoS), last accessed online 14.09.2011.
*Data sourced from Thomson Reuters Web of Knowledge (formerly referred to as ISI Web of Science).*

**4. Characteristics and discussion**

Together, both examples in the previous section illustrate how the effect of the partition approach can differ in strength and direction, depending on the particular subjects investigated and on the underlying normalization variant that it is combined with. Analysis of larger samples and several of these variants will be needed for a more clear and complete view on the possible amplitude and direction of effects. The paragraphs below briefly discuss the general characteristics of Partition-based Field Normalization relevant in a context of deciding whether or not to use the indicator in a particular analysis.

*Position among standard methodologies*:

Partition-based Field Normalization can be seen as an intermediary approach between standard field normalization and journal normalization, given the intermediate dimension of the set of journals determining the expected citation rates (Table 9).

**Table 9.** Global reference domains determining the expected citation rates for an observed publication record in different normalization variants.

| | Normalization variant | | |
|---|---|---|---|
| | **Journal Normalization** | **Partition-based Field Normalization** | **Standard Field Normalization** |
| **Global reference domain ($D$) containing all journals determining expected citation rates:** | $D_j$ = All journals containing publications from the publication record. | $D_p$ = All journals classified in exactly the same combination of subject categories as at least one journal that contains publications from the publication record. | $D_f$ = All journals classified in any of the subject categories in which at least one journal that contains publications from the publication record is classified. Journals classified in $N$ subject categories are counted partially (1/$N$) per subject category. |
| | $D_j \subseteq D_p \subseteq D_f$ | | |

*World average equal to one*:

When calculated for all subject categories combined, Partition-based Field Normalization yields a value of one, representing the world average, like standard field normalization does. Partition-based Field Normalization does the same for each discipline and partition cell separately, while standard field normalization does so per subject category in its 'reduced' form, i.e. counting only a fraction 1/$N$ of publications that are classified in $N$ subject categories.

*More closely fit reference domain*:

By its more closely fit global reference domain around an observed publication record, the potential advantages of Partition-based Field Normalization lie in particular in applications to highly specialized publication sets. These will contain publications from a relatively small volume of cells, leaving a relatively high volume of cells out of scope in Partition-based Field Normalization, as compared to standard methodology.

Standard field normalization may enlarge the reference domain well beyond the variation in publication records of researchers in a same research area, including contributions from unrelated research areas with very different citation rates. Because of this more 'loosely fit' reference domain, a lower confidence is attached to standard field normalized citation rates at the level of individual researchers, while this is a well established indicator at higher aggregation levels.

The partition approach makes an easy to implement step towards a more closely fit field delimitation that excludes influence from publications and evolutions in unrelated partition cells. At the same time, the closer position of the reference domain to the set of journals published in, enhances the importance to also include an indicator representing journal choice in the analysis, in particular in research areas where this choice may strongly vary.

*No intermediacy-assumption on intersections of subject categories*:

The content of a journal assigned to multiple subject categories is potentially of interest to researchers active in the 'core' of several of these subject categories, adding to the potential citing research community, and consequently enhancing the journal's impact factor and the expected citation rate to a level possibly exceeding the average over the subject categories involved. In standard methodology, the expected citation rate for publications in an intersection of subject categories is 'forced' between the expected citation rates for the separate intersecting subject categories. In Partition-based Field

Normalization, it can occur that the expected citation rate is higher in an intersection than in any of the intersecting subject categories.

Whether or not to apply Partition-based Field Normalization is only one of many methodological choices that may influence results of the field normalized citation rate, in particular at lower aggregation levels. Bibliometric indicators host a variety of evident and less visible choices, made for their adequacy in view of what needs to be measured, or in some cases for rather practical reasons. For optimal results, all choices need to be made consciously, evaluating which would fit the purpose of the analysis best through the aspects that it enhances or attenuates. Some recently debated choices related to field normalized citation rates are briefly listed below (recent discussions on a number of them can be found in: Bornmann, 2010; Leydesdorff and Opthof, 2010; Lundberg, 2007; Opthof and Leydesdorff, 2010; van Raan et al., 2010; Waltman et al., 2011a; Zitt et al., 2005; and references therein):
- The particular fixed subject category structure that is used as a basis for citation analysis, and its scale (fields, subfields, …).
- Global field normalization (in *CPP/FCSm* and *NMCS*) or field normalization per publication (in *MNCS*).
- Arithmetic averages or percentile impact classes as a standard for comparison of citation rates.
- Fractional or whole citation counting.
- Fractional or whole publication counting.
- Separate normalization per document type or not.

Without aiming to be exhaustive, this brief list illustrates that Partition-based Field Normalization is only one of many options that can be chosen in search for the best approach to a specific kind of subject and research question. Like Partition-based Field Normalization, many other choices are likely to affect results in particular for small publication records. One particular option and its effects are always situated and to be interpreted in the complex context of other choices and their effects.

**5. Conclusion**

*Partition-based Field Normalization* is an easy to implement adaptation of standard field normalization allowing a closer fit of the reference domain around an observed publication record. The partition approach can be combined with different existing field normalization variants that use pre-defined subject categories, regardless of whether citation rates are normalized globally or per publication, or whether expected citation rates are based on averages or percentile classes. Examples applying the partition approach to limited publication sets demonstrate that its influence on results strongly depends on the subjects investigated and on the other field normalization options it is combined with, and that its effect may be larger as well as smaller than that of other methodological choices.

By its more closely fit reference domain, the partition approach may offer an advantage in particular in an area where standard methodology until now could not be applied with the same confidence, i.e. applied to highly specialized research records, such as those of individual researchers or resulting from interdisciplinary research.

A clear view on the role that Partition-based Field Normalization can play in bibliometric analysis of particular kinds of subjects requires further investigation. This paper aims to facilitate this by clearly and accurately describing the partition approach using simple examples. Such studies should involve more complete publication records for larger samples of subjects than the simple examples shown in this paper for demonstration purposes and include comparisons to results from other methodologies such as peer ratings.

Besides field normalized citation rates, which are the focus of this paper, also other areas offer potential for application of a partition approach, handling bibliometric variables per partition cell. In the area of impact factors for instance, partition cells could form an alternative reference set of publications for the recently proposed Audience Factor (Zitt and Small, 2008). The various application possibilities are sure to offer interesting potential for further investigation.

**Acknowledgements**